\documentclass[12pt,preprint]{aastex}

\slugcomment{Version of 7 May 2007}
\shorttitle{Milagro upper limits on VHE emission from short GRBs}
\shortauthors{Atkins et al.}

\begin{document}

\title{Milagro Constraints on Very High Energy Emission from 
Short Duration Gamma-Ray Bursts}
\author{
A.~A.~Abdo,\altaffilmark{\ref{msu}}
B.~T.~Allen,\altaffilmark{\ref{uci}}
D.~Berley,\altaffilmark{\ref{umcp}} 
E.~Blaufuss,\altaffilmark{\ref{umcp}} 
S.~Casanova,\altaffilmark{\ref{lanl}}
B.~L.~Dingus,\altaffilmark{\ref{lanl}} 
R.~W.~Ellsworth,\altaffilmark{\ref{georgemason}} 
M.~M.~Gonzalez,\altaffilmark{\ref{unam}} 
J.~A.~Goodman,\altaffilmark{\ref{umcp}} 
E.~Hays,\altaffilmark{\ref{umcp},\ref{hayscurrent1},\ref{hayscurrent2}} 
C.~M.~Hoffman,\altaffilmark{\ref{lanl}}
B.~E.~Kolterman,\altaffilmark{\ref{nyu}}
C.~P.~Lansdell,\altaffilmark{\ref{umcp}}
J.~T.~Linnemann,\altaffilmark{\ref{msu}}
J.~E.~McEnery,\altaffilmark{\ref{mcenerycurrent}}
A.~I.~Mincer,\altaffilmark{\ref{nyu}} 
P.~Nemethy,\altaffilmark{\ref{nyu}} 
D.~Noyes,\altaffilmark{\ref{umcp}} 
J.~M.~Ryan,\altaffilmark{\ref{unh}} 
F.~W.~Samuelson,\altaffilmark{\ref{fda}} 
P.~M.~Saz~Parkinson,\altaffilmark{\ref{ucsc}}
A.~Shoup,\altaffilmark{\ref{osu}} 
G.~Sinnis,\altaffilmark{\ref{lanl}} 
A.~J.~Smith,\altaffilmark{\ref{umcp}} 
G.~W.~Sullivan,\altaffilmark{\ref{umcp}} 
V.~Vasileiou,\altaffilmark{\ref{umcp}}
G.~P.~Walker,\altaffilmark{\ref{lanl}}
D.~A.~Williams,\altaffilmark{\ref{ucsc}}
X.~W.~Xu\altaffilmark{\ref{lanl}} 
and 
G.~B.~Yodh\altaffilmark{\ref{uci}}} 

\altaffiltext{1}{\label{msu} Department of Physics and Astronomy, Michigan State University, 3245 BioMedical Physical Sciences Building, East Lansing, MI 48824}
\altaffiltext{2}{\label{uci} Department of Physics and Astronomy, University of California, Irvine, CA 92697}
\altaffiltext{3}{\label{umcp} Department of Physics, University of Maryland, College Park, MD 20742}
\altaffiltext{4}{\label{lanl} Group P-23, Los Alamos National Laboratory, P.O. Box 1663, Los Alamos, NM 87545}
\altaffiltext{5}{\label{georgemason} Department of Physics and Astronomy, George Mason University, 4400 University Drive, Fairfax, VA 22030}
\altaffiltext{6}{\label{unam} Instituto de Astronom\'{i}a, Universidad Nacional Aut\'{o}noma de M\'{e}xico, D.F., M\'{e}xico, 04510}
\altaffiltext{7}{\label{hayscurrent1} Current address: High Energy Physics Division, Argonne National Laboratory, Argonne, IL 60439}
\altaffiltext{8}{\label{hayscurrent2} Current address: Enrico Fermi Institute, University of Chicago, Chicago, IL, 60637}
\altaffiltext{9}{\label{mcenerycurrent} NASA Goddard Space Flight Center, Greenbelt, MD 20771}
\altaffiltext{10}{\label{nyu} Department of Physics, New York University, 4 Washington Place, New York, NY 10003}
\altaffiltext{11}{\label{unh} Department of Physics, University of New Hampshire, Morse Hall, Durham, NH 03824} 
\altaffiltext{12}{\label{fda} Office of Science and Engineering Laboratories, Center for Devices and Radiological Health, U.S. Food and Drug Administration.}
\altaffiltext{13}{\label{ucsc} Santa Cruz Institute for Particle Physics, University of California, 1156 High Street, Santa Cruz, CA 95064}
\altaffiltext{14}{\label{osu} Ohio State University, Lima, OH 45804}

\begin{abstract}

Recent rapid localizations of short, hard gamma-ray bursts (GRBs) by the Swift and HETE 
satellites have led to the observation of the first afterglows and the measurement of the first 
redshifts from this type of burst~\citep{2005Natur.437..845F,2005Natur.437..851G,
2005Natur.437..855V,2005Natur.438..988B, 2005Natur.438..994B}. 
Detection of $>$100 GeV counterparts would place powerful constraints on GRB mechanisms.
Seventeen short duration ($<$5 s) GRBs detected by satellites occurred within the field of view of 
the Milagro gamma-ray observatory between 2000 January and 2006 December. We have searched the 
Milagro data for $>$100 GeV counterparts to these GRBs and find no significant emission 
correlated with these bursts. 
Due to the absorption of high-energy gamma rays by the extragalactic background light (EBL), 
detections are only expected for redshifts less than $\sim$0.5. While most long 
duration GRBs occur at redshifts higher than 0.5, the opposite is thought to be true of short GRBs.
Lack of a detected VHE signal thus allows setting meaningful fluence limits. 
One GRB in the sample (050509b) has a likely association with a galaxy 
at a redshift of 0.225, while another (051103) has been tentatively linked to the nearby galaxy 
M81. Fluence limits are corrected for EBL absorption, either using the known measured 
redshift, or computing the corresponding absorption for a redshift of 0.1 and 0.5, as well as 
for the case of z=0.

\end{abstract}

\keywords{gamma rays: bursts --- gamma rays: observations} 

Gamma-ray bursts (GRBs) have long been classified by their 
durations into long and short bursts~\citep{1981Ap&SS..75...47M,1984Natur.308..434N}. 
Later classification schemes took into account the combination of both the temporal and spectral 
properties~\citep{1993ApJ...413L.101K} leading to what are currently known as 
short, hard bursts and long, soft bursts. Recent classification schemes list as 
many as ten different criteria to try and distinguish between these two 
populations~\citep{2006astro.ph..5570D}. The fraction of bursts that fall in 
each category is instrument-dependent, with BATSE finding approximately 25\% of bursts to be
``short''\citep{1999ApJS..122..465P}, while the equivalent fraction for Swift is closer to 
10\%~\citep{2006AIPC..838...14G}. The discovery of the first X-ray afterglow from a long duration 
GRB~\citep{1997Natur.387..783C} led to a rapid string of observations validating the fireball 
shock model of GRBs~\citep{1992MNRAS.258P..41R,1993ApJ...405..278M}, culminating in the 
observation of a GRB-supernova association~\citep{2003Natur.423..847H,2003ApJ...591L..17S} 
confirming that at least some GRBs are related to the deaths of massive stars, as predicted by 
the ``collapsar'' model~\citep{1993ApJ...405..273W}.

Until recently, however, all the observations of afterglows (and therefore, most of the 
understanding about GRBs) came from long duration GRBs. The first detection of the afterglow of
a short, hard burst -- for GRB 050509b~\citep{2005Natur.437..851G} -- was followed by 
others~\citep{2005Natur.437..845F,2005Natur.437..855V,2005Natur.438..988B}, and there are now 
approximately half a dozen measured redshifts for short, hard bursts~\citep{2006AIPC..838...25H}. 
Although some of these redshifts are less secure than others, their average ($\sim$ 0.3--0.5) 
is significantly lower than the typical redshift of long duration bursts. The location of several 
of these short bursts in old galaxies with little star formation, unlike the association of long 
GRBs with active star-forming regions, seems to rule out the collapsar model for these bursts and 
favors instead merger models involving binary neutron stars or black hole-neutron star systems as 
the progenitors for these bursts. One predicted consequence of these 
models~\citep{2006ApJ...650..998R} is that the neutron-rich outflows expected from these mergers 
would lead to pion decay photons at $\sim$60 GeV which could be detected by Milagro. 

The detection 
of gamma rays in the GeV-TeV regime is affected by the extragalactic infrared background light 
(EBL)~\citep{nikishov61}. The amount of gamma-ray absorption due to the EBL 
is not well determined, though it is a strong function of redshift and energy. One 
model~\citep{primack05}, recently validated by HESS observations~\citep{2006Natur.440.1018A}, 
predicts an optical depth of roughly unity to 500 GeV (10 TeV) gamma rays from a redshift of 
0.2 (0.05). The significantly lower redshift of short duration GRBs compared to long duration 
ones makes them particularly suitable candidates for very high energy (VHE) emission studies, such 
as possible with the Milagro detector. On the other hand, their much lower luminosity means their 
possible emission at higher energies is also expected to be substantially lower than the brighter, 
long duration bursts.

Previous searches for VHE emission from GRBs, both long and short, have produced no conclusive
detection to date. Milagrito, a prototype of Milagro, reported evidence for emission above 
650 GeV from GRB 970417a, with a (post-trials) probability of 1.5$\times10^{-3}$ of being a 
background fluctuation~\citep{atkins00a,atkins03}. More recent Milagro searches have yielded no
conclusive detection~\citep{2005ApJ...630..996A,2006astro.ph.11457S}.
Evidence at about the 3 sigma level from the 
HEGRA AIROBICC array has been published for emission above 20 TeV from GRB 920925c~\citep{padilla98}. 
Follow-up observations above 200 GeV by the Whipple atmospheric Cherenkov 
telescope~\citep{connaughton97,2007ApJ...655..396H} did not find any high energy afterglow from 
the GRBs observed. Recently, the MAGIC group have reported upper limits on the gamma-ray flux 
in the 85-1000 GeV energy range from the 9 GRBs\footnote[1]{Unfortunately, 4 out of the 9 GRBs that 
MAGIC observed had measured redshifts in excess of 3.5, making it virtually impossible for any VHE 
gamma rays to reach Earth.} they observed in their first year of operations, including the afterglow of the  
the short duration HETE burst 060121~\citep{2006astro.ph.12548A}. The MAGIC list includes 
GRB 050713a, for which they had the fastest response so far, beginning their observations 40 s 
after the burst onset~\citep{2006ApJ...641L...9A}. 
Because searches carried out with atmospheric Cherenkov telescopes, like MAGIC or Whipple, involve 
slewing a telescope to the right location in the 
sky and are limited by their relatively small fields of view and duty cycles, Milagro is the best suited 
instrument for observing the shortest GRBs at very high energies.

In this paper we place limits on the VHE emission from short duration\footnote[2]{The term ``short 
duration'' is used in the paper to refer only to the duration of the burst being less than 5 seconds, 
while the term ``short, hard'' burst is used in the usual more narrow sense found in the literature, 
based on the timing and spectral properties of the burst, as introduced by
~\cite{1993ApJ...413L.101K}.} GRBs which might help constrain models of their progenitors. We 
selected all known bursts detected by satellites which occurred
in the Milagro field of view and had a duration of 5 seconds or less. This duration was 
chosen, rather than 2 seconds, in part due to the recent work of \cite{2006astro.ph..5570D}, but 
also in order to be more inclusive. In the following 
section we describe the detector, Milagro, which was used to perform the search. We describe in 
some detail the new, low-energy-threshold trigger which was especially designed to increase 
Milagro's sensitivity to GRB detections. In section 3, the sample of short duration GRBs analyzed 
in the paper is presented, with a special emphasis on GRB 050509b, the most promising candidate 
in the sample. Section 4 describes the analysis carried out to search for emission, both prompt 
and delayed. Finally, in Section 5 we discuss the main results and summarize our conclusions.

\section{The Milagro Observatory}

Milagro is a TeV gamma-ray detector which uses the water Cherenkov technique to detect 
extensive air showers produced by VHE gamma rays as they traverse the Earth's 
atmosphere~\citep{atkins00b}. Milagro is located in the Jemez Mountains of northern 
New Mexico (35.9$^\circ$ N, 106.7$^\circ$ W) at an altitude of 2630 m above sea level,  
and has a field of view of $\sim$2 sr and a duty cycle of over 90\%, making it an 
ideal all-sky monitor of transient phenomena at very high energies, such as GRBs. 
The effective area and energy threshold of Milagro are a function of 
zenith angle, due to the increased atmospheric overburden at larger zenith 
angles, which tends to attenuate the particles in the air shower before they reach the ground. 
The sensitivity of Milagro varies slowly with zenith angle from 0 to $\sim$30 degrees and then 
decreases more rapidly~\citep{2005ApJ...630..996A}.

For the data sample used in this analysis, the typical single shower angular resolution is approximately 
0.7 degrees; however, at lower energies there are fewer photomultiplier tubes hit so the angular 
resolution is about 1 degree. The energy response of Milagro is rather broad 
with no clear point to define as an instrument threshold. To obtain a rough guide of the range of 
energies to which Milagro is sensitive, we consider a 
power-law spectrum with a differential photon index, $\alpha$, of -2.4. The energy (E$_{5}$) above 
which 95\% of the triggered events  from such a spectrum are obtained is approximately 350 GeV, 
the energy (E$_{95}$) below which 95\% of the triggered events occur is 30 TeV, and the median 
energy is 3 TeV. This illustrates the breadth of the energy response of Milagro, showing that 
the Milagro detector has significant sensitivity below energies of several hundred GeV. 

\begin{figure}[htbp]
\centering 
\rotatebox{0}{\resizebox{!}{12cm}{
\includegraphics{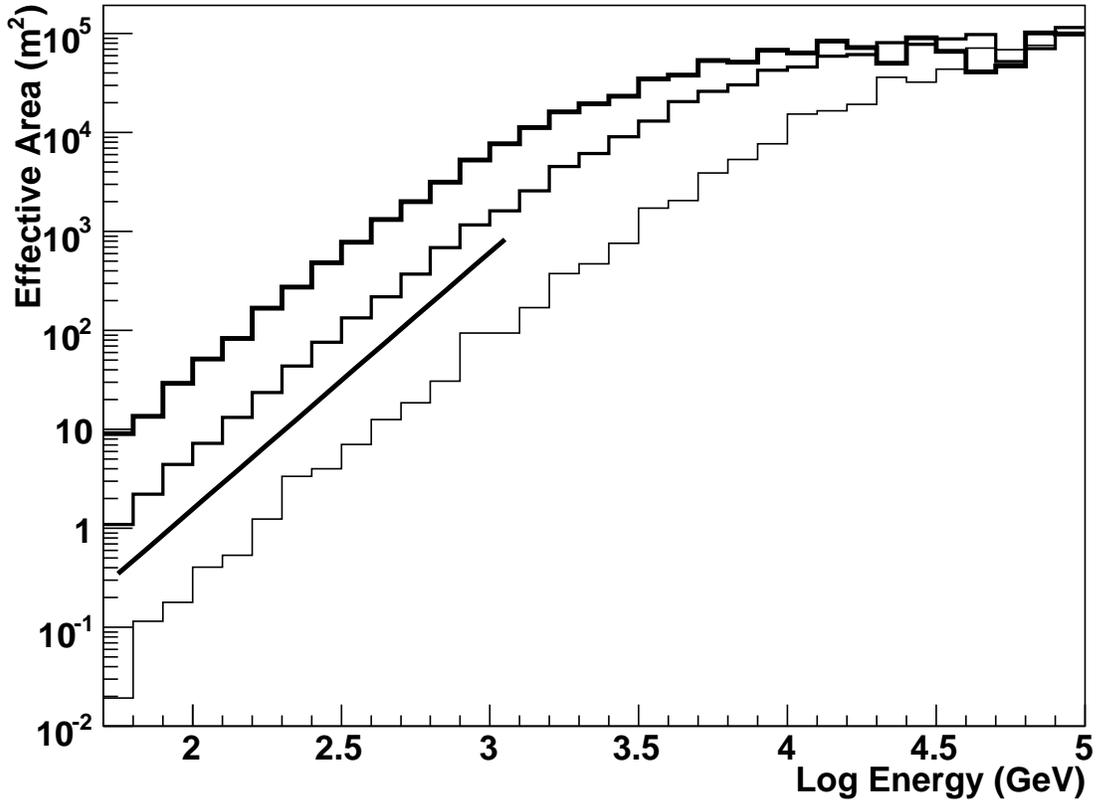}}}
\caption{Effective area of Milagro for gamma rays as a function of energy for three 
different zenith angles. The straight line is a power law $E^{2.6}$ (with arbitrary 
normalization). The different curves (in decreasing order of thickness) reflect the 
effective area for zenith angles of 10$^\circ$, 30$^\circ$, and 45$^\circ$ (roughly corresponding 
to GRBs 050509b, 050505, and 040924). The figure illustrates the decrease in effective area with 
zenith angle. The limited number of simulated showers at the highest energies results in fluctuations 
in the curves above 10$^4$ GeV.
\label{fig1}}   
\end{figure}

The Milagro sensitivity as a function of energy can be understood as a simple 
consequence of one dimensional cascade shower theory.
The fluctuations in the amount of energy reaching a certain detector level from a gamma-ray 
shower arise primarily because
of variations in the depth of the first interaction, which follows a probability
distribution $P\sim e^{-\frac{9}{7}X}$, where $X$ is the depth of the interaction in radiation lengths.
According to Approximation B~\citep{1941RvMP...13..240R}, after shower maximum ($>$10 km above 
sea level for the energy range of Milagro, well above the altitude of the Milagro detector), 
the average number of particles in a gamma-ray shower, as well 
as the amount of energy, decreases exponentially as shower particles 
are absorbed by the atmosphere. 
From the longitudinal shower profile obtained in Approximation B,
the number of radiation lengths deeper in the atmosphere, $X$, which a gamma-ray 
of energy $E$ must penetrate in order to deposit energy at the ground equivalent to that of a typical 
shower of higher energy $E_{thr}$ is given by $X\simeq2\,ln(E_{thr}/E)$. 
So the probability that a gamma ray shower of energy $E$ has a certain minimum amount of energy reaching 
the ground is given approximately by $P(E)\sim(\frac{E}{E_{thr}})^{2.6}$.
In other words, the low energy effective area scales like a power law in energy. Figure~\ref{fig1} 
shows that the effective area of Milagro does, indeed, follow this power law.
As seen from Figure~\ref{fig1}, the ratio of the effective area at 100 GeV vs 1 TeV is $\sim$0.005, 
roughly what is predicted by the previous formula. The effective area of Milagro at a median energy 
of $\sim$4 TeV has been confirmed by the measurement of the flux from the Crab, in agreement with 
atmospheric Cherenkov telescope measurements. For more details on Milagro see \cite{atkins03b}.

\begin{figure}[htbp]
\centering 
\rotatebox{0}{\resizebox{!}{12cm}{
\includegraphics{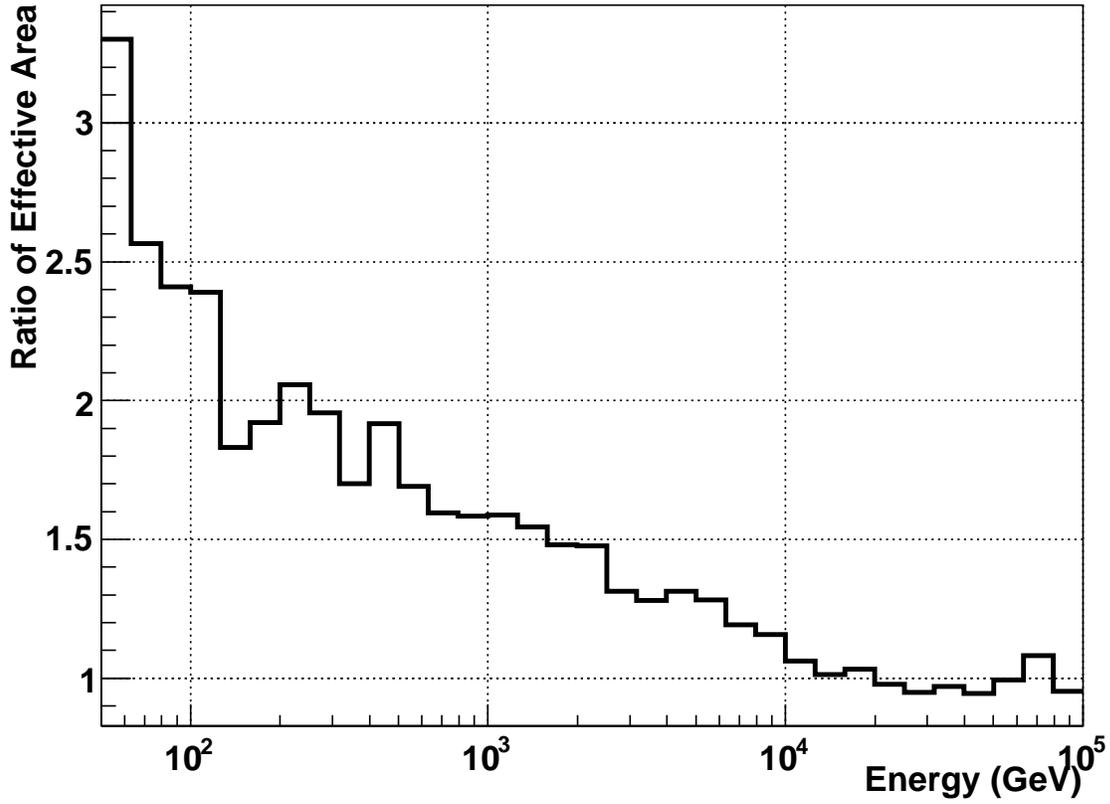}}}
\caption{Relative increase in effective area between the simple (55 tube) multiplicity trigger
and the VME programmable trigger, as applied to GRB 050509b. The figure shows an increase
in effective area using the new trigger of more than 50\% at 1 TeV and around 150\% at 100 GeV, 
relative to the old trigger.\label{fig2}}   
\end{figure}

During the period covered by these observations, the Milagro trigger underwent a significant 
upgrade. Until 2002, the Milagro trigger consisted of a simple multiplicity count of 
the number of photomultiplier tubes hit out of the 450 in the top layer of
the pond. This threshold was set to between 50 and 70 tubes hit within a 200 ns 
time window to maintain the trigger rate at $\sim$1400-1600 Hz, the maximum sustained rate that can 
be handled by the Milagro data acquisition system with a reasonable deadtime ($\sim$5\%)
\footnote{The deadtime to record single triggers depends instead on the digitization time, which
scales with the number of hit PMTs, and is $<$50 $\mu$s. Triggers separated by as little
as 30 $\mu$s are routinely recorded.}.
A lower trigger threshold would lower the energy 
threshold of Milagro, thus making it more sensitive to GRBs. Based on the knowledge 
that most of the increase in the rate as the multiplicity requirement is lowered comes from single 
muon events which produce enough light to trigger the instrument but cannot be fit to a 
shower plane, a new programmable trigger was custom-designed for Milagro.
It is known from Monte Carlo simulations that gamma-ray events can be
reconstructed with as few as 20 tubes hit. A high angle muon traveling
across the pond nearly horizontally produces light which arrives over
a longer time period than the shower particles, so by making a cut on the 
time development of the event, it is possible to eliminate these muon events. A custom 
VME trigger module was built, allowing the use of multiple 
trigger conditions and including the rise time of the pulse representing the number of
struck tubes in the top layer as one of the triggering criteria.  The new trigger 
greatly increased the number of low energy showers detected, while maintaining
a manageable overall trigger rate and dead time. Figure~\ref{fig1} shows the effective 
area of Milagro to gamma rays as a function of energy for three different zenith angles.
Figure~\ref{fig2} shows the significant increase in sensitivity gained from the new 
trigger, relative to the old simple multiplicity trigger, especially at energies below 
100 GeV, where detection of GRBs is most likely. The VME trigger was installed in January 
2002 and became fully operational on 19 March 2002. The column labeled ``Notes'' in 
Table~\ref{grb_table} identifies the bursts in our sample for which the VME trigger was in 
operation.

\section{The GRB sample}
There is no sharp cutoff point between long duration and short 
duration bursts; these two populations of GRBs have overlapping distributions in duration. 
Although earlier studies determined that an effective T90 (duration over which the cumulative 
counts over the background increase from 5\% to 95\% of the total) cut for separating short 
from long bursts should be approximately 2 seconds~\citep{1993ApJ...413L.101K}, more recent 
work~\citep{2006astro.ph..5570D} suggests that bursts shorter than {\em five} seconds 
have a higher probability of belonging to the short duration class than the long 
duration one, so we have chosen to include GRBs with 
durations up to 5 seconds in this list of ``short duration'' bursts.

In the 7 years since Milagro began operations (2000 January to 2006 December), there have 
been approximately 100 known GRBs detected by satellites which have been in the Milagro field 
of view. Of these, seventeen had measured durations of five seconds or shorter. Many of the bursts
in this study were detected by the Interplanetary Network 
(IPN\footnote[3]{See {\tt http://www.ssl.berkeley.edu/ipn3/}}), and their locations were not 
immediately known to experiments on the ground, making it very unlikely that a redshift 
could be determined. More recent bursts detected by Swift and HETE have benefited from 
extensive multi-wavelength observations from the ground and are therefore far better 
studied. One burst in our sample (GRB 001204) was obtained from the BeppoSAX 
GRBM catalog~\citep{guidorzi}. 

Table~\ref{grb_table} lists the sample of 17 bursts that we 
analyzed for this paper. Four of the bursts in the sample (000330, 000408, 000424, and 010104) 
were presented in an earlier paper summarizing the first two years of Milagro observations 
of GRBs~\citep{2005ApJ...630..996A} and are included here for completeness. One of these bursts 
(GRB 010104) has recently been found to have occurred at a significantly different location 
than previously thought~\citep{hurley}, so we take this opportunity to present our results on 
this burst at the new location. 
The first column of the table gives the GRB name, which, following the usual convention, 
represents the UTC date (YYMMDD) on which the burst took place. The second column gives the 
instrument(s) that detected the burst. We list the IPN as an instrument, although it consists of a 
network of many satellites, a different set of which may detect any given burst.
The third column gives the time of the burst, 
represented by the UTC second of the day. Column four gives the coordinates (right ascension 
and declination, in degrees) of the burst. All the bursts listed in the table except for one 
(GRB 000330) were localized to an error region significantly smaller than the Milagro angular resolution.
For GRB 000330, the position error was approximately 5 degrees, so the upper limit was computed using 
the most significant bin within that region, as described in~\cite{2005ApJ...630..996A}.
For one burst, GRB 000607, the coordinates are not known unambiguously; 
the IPN sometimes determines two possible error regions and in this case only one of them was in
the field of view of Milagro.
The fifth column gives the duration of the burst, as reported by the different instrument teams. 
Column six lists the zenith angle of the burst at Milagro, in degrees. We include only bursts for 
which the zenith angle was less than approximately 50$^{\circ}$.
The effective area of Milagro at zenith angles greater than 50$^{\circ}$ becomes negligible in 
the energy range where we expect GRB emission to be detectable (e.g. $<$ 1 TeV). Column seven 
gives the value of the redshift, if measured. 

For those bursts with no measured redshift, we take into account the effect of absorption in 
computing the upper limits by considering two different redshifts: z=0.5 and z=0.1. We also 
give limits for the case z=0 (i.e. nearby bursts).
By their very nature, short duration bursts are much more difficult to localize than long 
duration bursts. In addition to being very brief events, they also tend to be much less 
luminous than long duration GRBs, making it much more challenging to obtain
redshifts from these bursts than from long GRBs. GRB 040924, detected by 
HETE~\citep{Fenimore}, was the first short duration burst to have a measured 
redshift~\citep{wiersema}, although its spectrum was considered 
too soft to be part of the short, hard population and it has been speculated that it may belong 
to the short duration tail of the long duration GRB population~\citep{2005ApJ...628L..93H}. 
GRB 050509b was the first short, hard burst for which an afterglow was detected. As it is the 
most interesting burst in the sample, we describe it in more detail in the following paragraph. 
The remaining columns of Table~\ref{grb_table} present the Milagro results, which we describe 
later. 

The detection of an X-ray afterglow from GRB 050509b by Swift~\citep{2005Natur.437..851G}
represented the first time such an event had been observed from a short, hard burst. A
low probability ($\sim5\times10^{-3}$) of chance alignment suggests that this burst may be 
associated with a bright elliptical galaxy at a redshift of 0.225~\citep{2006ApJ...638..354B}. 
Subsequent detections of short, hard bursts~\citep{2005Natur.438..994B} have made this 
association more plausible and point to an origin of these bursts in 
regions of low star formation, thus disfavoring the collapsar model invoked for explaining the 
long duration bursts. At 10 degrees, the zenith angle of this burst is the most favorable 
in the list of 17 short bursts, and one of the most favorable of all bursts to have occurred in 
the Milagro field of view. Its redshift of 0.225 is the second or third lowest of those GRBs 
with known 
redshift in the Milagro field of view (depending on whether or not one believes GRB 051103 is 
associated with the nearby satellite galaxy M81), again, making it a very promising candidate. 
The 15--150 keV fluence of this burst, however, was measured by Swift to be 
(9.5 $\pm$ 2.5) $\times 10^{-9}$ erg cm$^{-2}$, making it one of the dimmest bursts 
detected by Swift~\citep{2005Natur.437..851G} and about forty times dimmer than the
next dimmest short duration burst in this sample. If the VHE emission of GRBs scales with the 
fluence measured at the lower energies, this would dampen significantly the expectations of 
detecting such emission from this burst.

\section{Data Analysis}

\begin{figure}[htbp]
\centering 
\rotatebox{0}{\resizebox{!}{12cm}{
\includegraphics{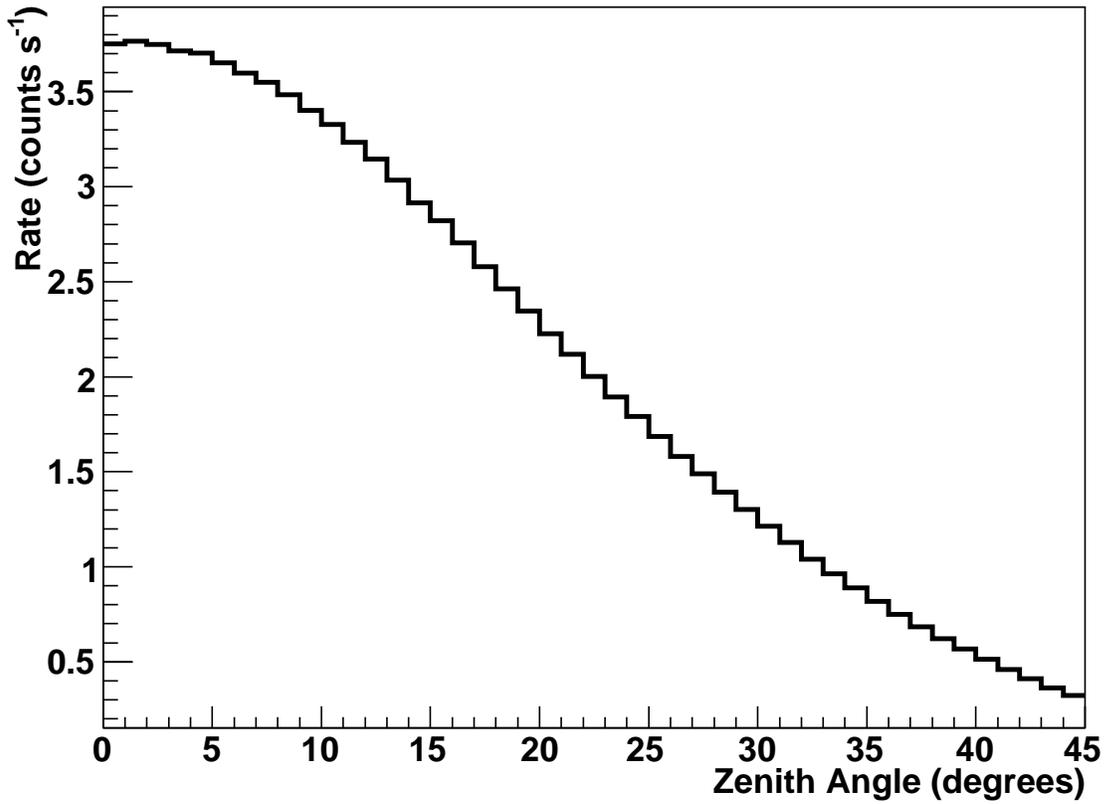}}}
\caption{Number of background events per second detected in a circular bin of radius 1.6 degrees, as a 
function of zenith angle. The background rate depends on the analysis cuts used as well as the detector 
configuration and atmospheric conditions on a particular day. The figure was made with data taken 
within one hour of GRB 050509b.\label{fig3}}   
\end{figure}

A search for an excess of events above those expected from the background was
made for each of the 17 bursts in the sample. The total number of events falling within 
a circular bin of radius $1.6^{\circ}$ at the location of the burst was summed for the 
duration of the burst. An estimate of the number of background events was then made by 
characterizing the angular distribution of the background using two hours of data 
surrounding the burst, as described in ~\cite{atkins03b}. Figure~\ref{fig3} shows the 
rate of background events detected by Milagro in a $1.6^{\circ}$ bin as a function of 
zenith angle. This background rate is a function of the trigger settings and the particular 
conditions of the detector on the given day and varies slightly from burst to burst. 
The significance of the excess (or deficit) for each burst was evaluated using 
equation [17] of~\cite{lima}. The 99\% confidence upper limits on the number of signal 
events detected, $\mathrm{N_{UL}}$, given the observed $\mathrm{N_{ON}}$ and the predicted 
background $\mathrm{N_{OFF}}$, is computed using the Feldman-Cousins 
prescription~\citep{feldman-cousins}. This upper limit on the number of gamma-ray events is then 
converted into an upper limit on the fluence. Using the effective area of Milagro, 
$A_{eff}$, and assuming a differential power-law photon spectrum, we integrate in the 
appropriate energy range and solve for the normalization constant. We chose a spectrum of the 
form $dN/dE=KE^{-2.4}$ photons/TeV/m$^2$. The spectrum of a GRB has never been measured 
above 100 GeV, so we must make an assumption of a suitable spectrum for
evaluating the limits. The average spectrum of the four brightest bursts observed by EGRET has a 
differential power law spectrum with index 1.95$\pm$0.25 over the energy range 30 MeV to 10 GeV, 
showing no sign of a cut-off, though only 4 gamma rays were detected above 1 GeV~\citep{dingus01}. 
The choice of 2.4 as the spectral index in the Milagro energy range allows for some softening of 
the spectrum at higher energy.

The normalization factor $K$ can be calculated by solving the equation 
$N_{UL}=\int{A_{eff}(dN/dE)e^{-\tau_{EBL}}dE}$, where $\tau_{EBL}$ represents the optical depth 
due to the EBL. Finally, we integrate the photon spectrum 
multiplied by the energy to obtain the corresponding value for the total fluence: 
$F=\int{E(dN/dE)dE}$, integrating from 0.05 to 5 TeV. For bursts of known 
(albeit uncertain) redshift (040924, 050509b, 051103, and 051221a), we use the 
optical depths predicted by~\cite{primack05} and take these into account in computing the preceding 
integrals, thus obtaining a more realistic upper limit which factors in the correct absorption 
due to the EBL. For the remaining bursts, we compute the upper limits assuming three possible 
values of the redshift: 0.5, 0.1, and 0.0.

In addition to searching for prompt emission from these bursts, we also searched for extended 
emission over a period of 312 seconds following the reported trigger time. This timescale is 
motivated both by the observations of late-time (several hundred seconds after the GRB trigger) 
X-ray flares during some GRB afterglows~\citep{2006ApJ...641.1010F,2005Sci...309.1833B}, 
which are predicted by some to emit in the GeV-TeV regime (e.g.~\cite{2006ApJ...641L..89W}), as well as 
by the discovery of a second higher energy component in GRB 941017. While the T90 for that burst 
was 77 s, the second, higher energy component (which has a fluence more than three times greater 
than the fluence in the BATSE energy range alone) had a duration of approximately 211 
seconds~\citep{gonzalez03}. 

\section{Results and Discussion}

None of the bursts in the sample showed significant VHE emission, either prompt or 
delayed. Column nine of Table~\ref{grb_table} gives the 99\% upper limits on the fluence, computed 
as described in the previous section over the duration (given in column five) of the burst. For 
comparison, we give the measured fluence in the keV band in column eight. Most models of 
VHE emission predict it should be correlated to the lower energy emission. In column 
ten, we give the 99\% upper limits on the fluence computed over a duration of 
312 seconds from the trigger time. 

The localization of several short, hard bursts to old, low-star-forming galaxies has led
to the speculation that their origins may be related to binary 
mergers, possibly double neutron star systems or black-hole neutron star binaries. 
\cite{2006ApJ...650..998R} propose that in such a scenario, the accretion of neutron star material 
would lead to the emission of a neutron-rich jet, which would emit $\pi^0$ decay photons in the 
100 GeV range. Several parameters and assumptions are important in this model, including the 
total isotropic-equivalent energy outflow of the burst, the total energy to mass flow ratio, 
$\eta$, and the initial neutron to proton number density ratio, $\xi_0$. 

Of the bursts considered in the sample, GRB 050509b is the most promising candidate, given its
known low redshift and its optimal zenith angle at Milagro. The attenuation due to the IR 
background in this case is not very significant. Using the \cite{primack05} model, the 
corresponding optical depth for the resulting 60 GeV photons at z=0.225 would be $\sim$0.04, 
leading to an attenuation of less than 5\%. Using the \cite{2006ApJ...650..998R} model with their 
standard parameters, $\eta$=316, and $\xi_0$=10, and using the measured isotropic luminosity in 
gamma rays, $E_{iso}$, the predicted flux from this GRB would be 
2.3$\times10^{-7}$ cm$^{-2}$s$^{-1}$ gamma rays of energy $\sim$60 GeV~\citep{razzaque}. The 
effective area of Milagro is approximately 90,000 cm$^2$ at 60 GeV for the given zenith angle 
of this burst, yielding approximately 0.02 per second, or less than 3$\times10^{-3}$ events for 
the duration (0.128s) of the burst, making this burst clearly undetectable. The next best 
candidate is GRB 061210. Despite having a much larger $E_{iso}$ than 050509b (about twenty 
times larger), this burst, assuming a redshift of 0.41 has a predicted flux of pion-decay photons 
comparable to 050509b~\cite{razzaque}. Given the less favorable zenith angle of this burst 
and the fact that the VME trigger was not operating at the time this burst took place, the effective 
area of Milagro for these events is approximately an order of magnitude lower than for the case of 
050509b. As discussed below, GRB 051103 might have been an SGR outburst in M81. If it is not 
an SGR outburst, but a binary merger at very low redshift, the model by \cite{2006ApJ...650..998R} 
would predict a significant detection of this burst in Milagro, had it 
occurred at a zenith angle $\leq$20 degrees, instead of at 50 degrees. This is despite having
a very low $E_{iso}$, more than an order of magnitude less than GRB 050509b.


It has been suggested that a fraction of short duration GRBs could be due to soft gamma-ray 
repeaters (SGRs) in nearby galaxies. There is some debate as to the exact fraction such objects
could represent, with estimates ranging from more than 1\%~\citep{ofek07} of short GRBs, to less 
than 40\%~\citep{lazzati}. We have presented upper limits at three different redshifts, including 
the case of z=0 which would be appropriate for bursts happening nearby. Indeed, the bright 
GRB 051103 detected by the IPN has been found to be consistent with an SGR flare originating in 
the nearby M81 galaxy group~\citep{ofek06}. Assuming this to be the location of the burst, we 
obtain a Milagro TeV upper limit (1.9$\times10^{-5}$erg cm$^{-2}$) which is lower than the 
IPN measured fluence of 2.3$\times10^{-5}$erg cm$^{-2}$. 

In conclusion, we have searched the Milagro data for prompt and delayed GeV--TeV emission from a 
collection of seventeen short duration ($<$ 5 s) GRBs which occurred in Milagro's field of view 
in the seven years since Milagro began operations in 2000. This represents the most comprehensive
search for very high energy emission from short GRBs ever performed. Due to the short duration and 
low rate of short bursts, such observations must carried out by an experiment like Milagro with its
large field of view of $\sim$2 sr and high duty cycle. While no emission was detected from any of these
short bursts, HAWC~\citep{HAWC},a next-generation version of Milagro, would have more than 15 times 
the sensitivity.  The GLAST Gamma-ray Burst Monitor with its BATSE like field of view of over 
2$\pi$ sr will detect many bright, short GRBs and simultaneous observations of the GLAST Large 
Area Telescope and HAWC will provide prompt spectra from keV-TeV energies to further our 
understanding of short GRBs.

\acknowledgements
We are grateful to Kevin Hurley for providing us with the details of the IPN bursts and 
for useful discussions regarding the use of such data. We thank Cristiano Guidorzi for discussions
regarding BeppoSAX data and Soeb Razzaque for help in comparing our data to his model. We are 
also grateful to James Bullock for sending us optical depth data from the \cite{primack05}
EBL model. We have used GCN Notices to select raw data for archiving and use in this search, and
we are grateful for the hard work of the GCN team, especially Scott Barthelmy. We acknowledge 
Scott Delay and Michael Schneider for their dedicated efforts in the construction and 
maintenance of the Milagro experiment.  This work has been supported by the National Science 
Foundation (under grants 
PHY-0245234, 
-0302000, 
-0400424, 
-0504201, 
-0601080, 
and
ATM-0002744) 
the US Department of Energy (Office of High-Energy Physics and 
Office of Nuclear Physics), Los Alamos National Laboratory, the University of
California, and the Institute of Geophysics and Planetary Physics.

\def \atel {The Astronomer's Telegram}
\def \apj {ApJ}
\def \aj {AJ}
\def \apjl {ApJL}
\def \mnras {MNRAS}
\def \iaucirc {IAUCIRC}
\def \em { }
\def \aap {A\&A}
\def \nat {Nature}
\def \araa {Anual Review of Astronomy and Astrophysics}


\clearpage
\thispagestyle{empty}
\begin{deluxetable}{lllllllllll}
\rotate
\renewcommand{\footnoterule}{\rule{0pt}{0pt}}
\tabletypesize{\tiny}
\tablewidth{0pt}
\tablecaption{List of short duration GRBs in the field of view of Milagro\label{grb_table}}

\tablehead{
\colhead{GRB} & 
\colhead{Instrument} &
\colhead{Time\tablenotemark{a}} &
\colhead{RA,Dec} & 
\colhead{T90/Dur.} & 
\colhead{$\theta$\tablenotemark{b}} & 
\colhead{z\tablenotemark{c}}	& 
\colhead{keV fluence\tablenotemark{d}} &
\colhead{TeV fluence UL\tablenotemark{e}} &
\colhead{312s TeV fluence UL\tablenotemark{f}} &
\colhead{Notes}
}

\startdata

000220  & BATSE & 17083.78 & 182.0,+66.0	& 2.4	& 48.8 R  &\nodata& 3.7e-7 (25--300) & 4.1e-3/1.5e-4/1.8e-5 & 1.8e-2/6.6e-4/7.9e-5 & T90$>$2s. High zenith angle.  \\
000330  & BATSE & 75449.40 & 358.3,+39.3\tablenotemark{*} 	& 0.2	& 30.0 S  &\nodata&\nodata 	& 3.0e-5/2.1e-6/7.5e-7 & 1.6e-4/1.1e-5/4.0e-6 & \\
000408	& BATSE,IPN & 9348.43& 137.3,+66.6	& 2.5	& 31.1 R  &\nodata&7.4e-6 (25--100) &2.7e-5/2.1e-6/7.2e-7 & 1.8e-4/1.4e-5/4.8e-6 & T90$>$2s. \\
000424  & BATSE	& 32666.36& 233.1,+71.8	& 5.0	& 36.2 S  &\nodata& 1.3e-6 (25--300)&6.4e-5/4.7e-6/1.4e-6 & 1.9e-4/1.4e-5/4.2e-6  & T90$>$2s. \\
000607	& IPN 	& 8690.4& 224.7,+13.5\tablenotemark{**}	& 0.12	& 41.8 R  &\nodata&5.3e-6 (15--5000) & 7.6e-5/4.1e-6/1.1e-6 & 5.6e-4/3.0e-5/8.4e-6 & One of two error regions.\\
001204  & BeppoSAX,IPN & 28870.25& 40.3,+12.9	& 0.25	& 47.8 S  &\nodata&3.7e-7 (25--100) & 1.8e-3/1.6e-4/2.0e-5 & 1.0e-2/8.9e-4/1.1e-4 & High zenith angle.\\
010104	& IPN 	& 62490.327 & 317.4,+63.5	& 2.0	& 44.8 R  &\nodata&4.3e-7 (25--100) &6.6e-5/3.5e-6/9.9e-7 & 5.8e-4/3.1e-5/8.7e-6 & Revised location\\
031026	& IPN 	& 5189.02 & 338.8,+0.02	& 0.24	& 45.3 R  &\nodata&\nodata	&1.1e-4/7.6e-6/2.0e-6 & 7.6e-4/5.3e-5/1.4e-5 & High zenith angle. VME trigger.\\		
040924  & HETE 	& 42731.36 & 31.6,+16.0  	& 0.6 	& 43.3 S  &0.859 &4.2e-6 (7--400)& 1.4e-3 &2.1e-2 & VME trigger. \\
050124  & Swift & 41402.87& 192.9,+13.0	& 4.1 	& 23.0 R  &\nodata&2.1e-6 (15--350) &1.3e-5/9.0e-7/3.1e-7 &1.2e-4/8.4e-6/2.9e-6 & T90$>$2s. VME trigger.\\
050509b & Swift & 14419.23 & 189.1,+29.0  	& 0.128 & 10.0 R  &0.225?&9.5e-9 (15--350)& 9.6e-7 & 2.1e-5 & VME trigger\\
051103  & IPN 	& 33942.186& 148.1,68.8  	& 0.17 	& 49.9 R  &0.0?&2.3e-5 (20--2000)& 1.9e-5 & 9.2e-5 & High zenith angle. VME trigger.\\
051221a & Swift,Suzaku & 6675.61 & 328.7,+16.9  	& 1.4 	& 41.8 S  &0.5465&3.2e-6 (20--2000) & 1.3e-4 &8.4e-4  & VME trigger\\
060210 	& Swift & 17929.8 & 57.7,+27.0 	& 5	& 43.4 S  &3.91  &7.7e-6 (15--150)  & \nodata  & \nodata& T90$>$2s. High z. VME trigger.\\
060313  & Swift & 726.29   & 66.6,-10.9 	& 0.8 	& 46.7 S  &\nodata &7e-5 (20--2000)& 1.4e-3/2.1e-4/1.9e-5 & 9.9e-3/1.5e-3/1.4e-4 & High zenith angle. VME trigger.\\
060427b & IPN 	& 85915.32 & 98.5,+21.3 	& 0.22	& 16.4 S  &\nodata &5.0e-6 (20--2000) & 1.8e-5/1.1e-6/3.6e-7 & 1.3e-4/7.6e-6/2.6e-6 & \\
061210	& Swift,Suzaku & 44439.33 & 144.5,+15.6 	& 0.8*** & 23.4 S  &0.41? &3.0e-7 (15--150)*** & 8.6e-6 & 1.7e-4 & \\

\enddata

\tablenotetext{a}{Time of burst, UTC second of the day}
\tablenotetext{b}{Zenith angle, in degrees; R=rising, S=setting}
\tablenotetext{c}{Redshift. A redshift of 0.5, 0.1, or 0 is assumed for those bursts where it is unknown.}
\tablenotetext{d}{Measured fluence in the keV energy range (given in parentheses), in erg cm$^{-2}$.}
\tablenotetext{e}{99\% upper limit on the fluence (0.05--5 TeV), in erg cm$^{-2}$ for the GRB duration, 
using the \cite{primack05} EBL absorption model. When no redshift is given in the table, the limits 
are calculated assuming three different redshifts: z=0.5/z=0.1/z=0.0}
\tablenotetext{f}{99\% upper limit on the fluence (0.05--5 TeV), in erg cm$^{-2}$ over a duration of 312s 
from the burst trigger. The same assumptions as in the previous column apply.}
\tablenotetext{*}{This GRB is the only one from this sample whose error region is larger than the Milagro bin size. See~\cite{2005ApJ...630..996A}.}.
\tablenotetext{**}{This location represents one of two possible error regions (the other is outside the field of view of Milagro).}
\tablenotetext{***}{These quantities apply only to the initial hard spike, not the entire burst.}

\end{deluxetable}

\clearpage

\end{document}